\documentclass[%
aip,
amsmath,amssymb,
reprint,%
]{revtex4-1}

\usepackage{graphicx}% Include figure files
\usepackage{dcolumn}% Align table columns on decimal point
\usepackage{bm}% bold math
\usepackage[utf8]{inputenc}
\usepackage[T1]{fontenc}
\usepackage{mathptmx}
\usepackage{hyperref}

\usepackage{color,soul}
\newcommand{\cblack}{\color{black} }

\graphicspath{{./images_fin/}}
\begin{document}
	
	\preprint{AIP/123-QED}
	
	% Force line breaks with \\
	\title{Particle modeling of the spreading of Coronavirus Disease (COVID-19)}
	
	\author{Hilla\ De-Leon}
	\email[E-mail:~]{hdeleon@ectstar.eu}
	\affiliation{INFN-TIFPA Trento Institute of Fundamental Physics and Applications, Via Sommarive, 14, 38123 Povo TN, Italy}
	\affiliation{
		European Centre for Theoretical Studies in Nuclear Physics and Related Areas (ECT*),
		Strada delle Tabarelle 286, I-38123 Villazzano (TN), Italy}
	\author{Francesco Pederiva}
	\email[E-mail:~]{francesco.pederiva@unitn.it}
	\affiliation{INFN-TIFPA Trento Institute of Fundamental Physics and Applications, Via Sommarive, 14, 38123 Povo TN, Italy}
	\affiliation{Dipartimento di Fisica, University of Trento, via Sommarive 14, I–38123, Povo, Trento, Italy}
	
	\date{\today}% It is always \today, today,
	% but any date may be explicitly specified
	
	\begin{abstract}
		By the end of July 2020, the COVID-19 pandemic had infected more than seventeen million people and had spread to almost all countries worldwide. In response, many countries all over the world have used different methods to reduce the infection rate, such as including case isolation, the closure of schools and universities, banning public events, and mostly forcing social distancing, including local and national lockdowns. We use a Monte-Carlo (MC) based algorithm to predict the virus infection rate for different population densities using the most recent epidemic data in our work. We test the spread of the Coronavirus using three different lockdown models, and eight various combinations of constraints, which allow us to examine the efficiency of each model and constraint. In this paper, we have tested three different time-cyclic patterns of no-restrictions/lockdown patterns. This model's main prediction is that a cyclic schedule of no-restrictions/lockdown that contains at least ten days of lockdown for each time cycle can help control the virus infection. In particular, this model reduces the infection rate when accompanied by social distancing and complete isolation of symptomatic patients.\end{abstract}
	
	\maketitle
	
	\section{Introduction}
	
	%\section{}\label{MC}

	Statistical mechanics provides a set of very powerful tools to model various biological and medical problems (see, for example, Refs.~\cite{blossey2019computational,buldyrev1998analysis,stanley1994statistical} and many more). One of the most studied these days is pandemics' diffusion, prompted by the current COVID-19 emergency (see, for example, Ref.~\cite{SOHRABI202071}). In addition, many current researches study the physical aspects of the spreading of the virus (see, for example, Refs. \cite{doi:10.1063/5.0012009,doi:10.1063/5.0011960,doi:10.1056/NEJMc2004973,doi:10.1063/5.0015984,doi:10.1063/5.0015984}). 
	Many techniques currently employed are based on the solution of differential equations (see, for example, Refs.~\cite{karin2020adaptive,prem2020effect,zhao2020modeling} and many more) or fitting formulae (see, for example, Refs.~\cite{bliznashki2020bayesian,olsson2020ongoing}). Both techniques are based upon varied parameters to obtain several scenarios that are then treated as parts of a statistical ensemble to analyze. For example, in many countries (e.g., Germany, Italy), there is an ample discussion about the role of the so-called $R_0$ parameter, i.e., the average number of individuals that a single actively infectious person can pass the virus to. The procedures to estimate $R_0$ are all based on a-posteriori analyses, but are usually part of the parameters that governments use to decide on measures to be taken.
	
	In this paper, we propose a method that is essentially based on modeling a population as a set of interacting classical particles, each one with three states relative to the health status (susceptible of infection, infected and contagious, and recovered/died), in which standard thermodynamical parameters (temperature, density) are used to describe the characteristics of the population. This method allows us to apply the model to very different situations, ranging from a city suburb population to a single university classroom. 
	The algorithm is based on standard Monte-Carlo procedures of sampling the transition among subsequent states, which are essentially sampled from a statistical distribution, in the spirit of transport MC algorithms.

	In our model, a healthy person (i) can become sick with a daily probability,
	$P_i=\sum_jP_{ij}$, where $P_{ij}$ is a function of the distance between each infected person $(j)$ in the area and the healthy person $(i)$. We are treating the Coronvirus spread as a "one-way" Ising-model Monte-Carlo as the following: A healthy person becomes sick as a result of an interaction with a sick person (or people), but a sick person stops being sick (i.e., recovered or died) within an average time of $\sim 14$ up to 40 days for the severs cases (see Ref.\cite{10.7554/eLife.57309} for the epidemiology data).
	After that time, the recovered person can no longer infect another person and cannot be sick again.

	In contrast to other infection models, such as SIR \cite{karin2020adaptive,prem2020effect,zhao2020modeling} in this approach, the parameter $R_0$, is a direct outcome of the simulation and not pre-assumed. This is achieved by the relation between $R_0$ and the doubling time $T_d$, which is a direct result of the infection probability chosen, which is, in turn, a function of observable epidemiological data and features of the studied population (e.g., average density on a given area or mobility). In the following, we will present the results of several simulations, meant to reproduce the spread of the Coronavirus in the presence of different lockdown constraints. The model's high flexibility enables us to control many parameters such as social distancing, infection from an unknown sourced, etc.
	In Section II, we will describe some details of the model. In Section III, the different models of lockdown considered will be discussed. Section IV is devoted to the presentation and discussion of the results, and Sec. V to the Conclusions.
	
	\section{The parameters and preliminary assumptions}
	
	Given the scarce information available, we had to make some assumptions based on current data, which may be more stringent than it might be required by the real nature of the virus. In particular, we based on Ref.~\cite{10.7554/eLife.57309} for the Coronavirus epidemiology data and on Refs.~\cite{doi:10.1063/5.0012009,doi:10.1063/5.0011960} for the physical properties of the virus. We are aware that this model cannot take into account every single spreading event, but since such events affect the initial $T_d$, (which is a function of the population density), and since the infection process is random, we expect that the existence of such events will be reflected in the numerical results. Also, in contrast to the real-life, here, there is no time gap between the infection and being tested positive for Coronavirus. Therefore, an immediate decrease in the rate of infection resulting from lockdown is expected in the model in contrast to the real data. \cite{endcoronavirus}.
	
	All simulations are performed, assuming a surface area unit of 1 km$^2$. Periodic boundary conditions are used, allowing to get rid of broad confinement effects (e.g., the lockdown of an entire province or city) and look at the local dynamics of the infections within that area. \cblack In this work, we modeled the spread of the COVID-19 virus as a function of the population density in a specific surface. The application of periodic boundary conditions means that we have an infinite number of identical systems; each system is a replica of the others. I.e., if a person leaves the simulation surface on one side, an identical person will enter to the surface from the other side. \cblack The population density is a function of the number of households in a certain area since it is crucial to distinguish between the infection among household and non-household contacts\cite{10.1093/cid/ciaa450}.
	%The population density has then to be interpreted as an "effective" density. For example, a nursing home with a common dining area is denser than the ones where all the residents eat in their rooms, and a neighborhood with many young families (which meet in a playground, near schools, etc.) is denser than the same neighborhood with less social activities. 
	
	%accompanied by a lower $R_0$ than a `gas" population density (a light population with high numbers of social activities - hot).
	
	\subsection{Parameterization of the model}
	We have identified a list of parameters and corresponding values that describe the population and the infection's kinetics. This list is obviously partial, but it could be quite easily extended.
	\begin{enumerate}
		\item The probability of developing symptoms over time $t$. This is described by a Gaussian peaked at $\bar{t}=5$ days and with standard deviation $\sigma_t=1$ day.
		\item The number of effective households, denoted by N.
		\item The fraction of "silent carriers," which have no symptoms (AKA asymptomatic) but can infect other people. Their fraction in the population is denoted by: $a_{silent}$, and the probability of transmitting the infection has been set to 0.5.
		\item Each sick person is considered contagious between the 3$^\text{rd}$ and the 7$^\text{th}$ day.
	\end{enumerate}
	Simulations are started with a single infected person (the \textit{zero patient}). In some runs, infections from an unknown source (a healthy person becomes sick without interaction with a known sick person) are allowed. For the first 14 days, the infection has not been detected yet, and the population walks freely without any restrictions. 
	In some of the simulations, we "force" sick people with symptoms to keep 8 meters (i.e., stay at home) after day 14. This restriction reduces the probability of the non-household infection.
	
	\subsection{Population dynamics}
	Our model is based on the principles of Brownian motion, such that for each day the population position ($R$) and displacement ($\Delta R)$ are given by:
	\begin{equation}
	R\rightarrow R+\Delta R~,
	\end{equation}
	where $\Delta R=\sqrt{\Delta x^2+\Delta y^2}$ is distributed normally:
	\[
	P[\Delta R] =\frac{1}{2\pi^2\sigma^2_{R}} \exp\left(-\frac{\Delta x^2+\Delta y^2}{2\sigma_{R}^2}\right)~,
	\]
	where $\Delta x$ ($\Delta y$) is displacement in the x-(y-) direction and $\sigma_{R}^2$, the variance, is a function of the diffusion constant, $D$:
	\begin{equation}
	\sigma^2_{R}=2Dt~,
	\end{equation}
	where t=1 day. For a Brownian motion the diffusion coefficient, $D$, would be related to the temperature, $T$, using the Einstein relation:
	\begin{equation}
	D=\mu k_{\text{B}}T~,
	\end{equation}
	where $\mu$ is defined as the mobility, $k_{\text{B}}$ is Boltzmann's constant
	and $T$ is the absolute temperature. By fixing $T=1$, the diffusion coefficient would be directly related to the mobility. Still, it is interesting to notice that the mobility could, in principle, be directly interpreted as a sort of thermal parameter. In our model, the time period when the population is allowed to move without restrictions is characterized by a large value of $\sigma_{R}$, (namely $\sigma_{\text{high}}=500$ meter), i.e. high temperature, while a lockdown is characterized by a lower $\sigma_{R}$, (here $\sigma_{\text{low}}=$%\textcolor{red}
	{$.5\times 10^{0.5}$ meter}, %THIS IS 1.5 meters... are you sure?)},
	i.e., low temperature. Thus, one can consider the infection rate problem in terms of heating/cooling of the system.
	
	\subsection{The infection probability}
	The core of the model, which contains most of the epidemiological data, is the probability for the $i^{\text{th}}$ healthy person to become sick. We assume that for each contact with another infected person, this process can be described by a Gaussian function of the distance, weighted with a factor that parametrizes the sick person’s conditions and social interaction. \footnote{Since the infection probability is a function of the absolute value of the distance between two people, and the standard deviation, several distributions such as a Gaussian distribution and a Lorentzian distribution could serve for modeling the infection probability under the assumption that the simulation’s dynamics dependents on $\sigma_r$ and not on the distribution’s tail. The choice of Gaussian distribution was since this is the typical distribution for thermal systems approaches for equilibrium, e.g., Maxwell Boltzmann distribution.}
	
	\begin{equation}
	\begin{split}
	P_i=&int\left(\sum_{j=1}^{n_{sick}}P_{ij} +\xi\right)=\\
	&int\left\{\sum_{j=1}^{n_{sick}}\exp\left[\frac{\left(r_i-r_j\right)^2}{2\sigma_r^2}\right]\times f\left(a_{silent},n_{out}\right) +\xi\right\}
	\end{split}
	\end{equation} 
	
	where:
	\begin{itemize}
		\item $r_i \left(x_i,y_i\right)$ is the location of the $i^{\text{th}}$ healthy person and $r_j \left(x_j,y_j\right)$ is the location of the $j^{\text{th}}$ sick person, so $|r_i-r_j|$ is the distance between them.
		\item $n_{sick}$ is the total number of sick people in the area. 
		\item $\sigma_r$ is the standard deviation (here $\sigma_r=2.4$ meters) %\textcolor{red}{this is very debated, it depends on whether you wear a ask or not, etc. We should add some appropriate citation}
		{since recent studies show that even a slight breeze can drive droplets arising from a human cough over more than 6 meters
			~\cite{doi:10.1063/5.0011960}}.
		\item $f(a_{silent},n_{out} )$ is a function that considers the social activity of the sick person and whether he has symptoms, which affects the spread of the virus outside the house. In our model, we estimate that infection by asymptomatic people is approximately 50\% lower than patients with symptoms.
		\item $\xi$ is a random number between 0 and 1, which allows us to consider some violation of the lockdown and the fact that even during a full lockdown, people continue going out of their homes to buy groceries, take a walk, etc.
	\end{itemize}
	Also, we are assuming that each sick person will infect some of his household members. Since the latest estimates are that household infections are $\sim 15\%$ from the known cases (without lockdown, \cite{jing2020household}), we estimated that the number of household infections is are of uniformly distributed between 0 and 3. This number is constant for all the simulations and does not depend on the population density, or the lockdown constrain.

	\section{Lockdown strategies}
	To date, lockdown is still being imposed in many countries to reduce $R_0$ and, as a result, to increase $T_d$, the doubling contagion time. It still remains an option if the disease starts spreading again without control, or if new local outbreaks should appear. In our work, we tested three different types of lockdown strategies. This choice is just representative of a potentially much broader set of options that could be analyzed using this method simply varying the corresponding parameters, and with a minimal computational cost. 
	For all the models presented in this paper we assumed the following conditions:
	\begin{itemize}
		\item Days 1-14: no restrictions.
		\item Days 15-50: full lockdown with moderate social distancing (SD), i.e., people are forced to keep 3 meters from each other.
		\item Days 51-200: people must wear face-masks so that the daily infection probability (for the non-household members) is reduced to:
		\begin{equation}
		int\left\{\sum_{j=1}^{n_{sick}}0.7\times\exp\left[\frac{\left(r_i-r_j\right)^2}{2\sigma_r^2}\right]\times f\left(a_{silent},n_{out}\right) +\xi\right\}
		\end{equation}
		
		{A very recent HKU hamster research shows that by wearing a proper mask, the infection probability can be reduced by a factor of 3 \cite{19_2020}. Therefore, given that not all the population wears a mask properly, and the findings of Ref. \cite{doi:10.1063/5.0015044}, we estimated the probability of infection when wearing masks 1.4 times lower than without masks.}
	\end{itemize}
	Days 1-14 are the heating phase of the system. Thus, we expect the fastest increase in the number of patients these days. For days 15-50, we predict a phase transition from a hot system to a colder system (almost solid-like), which will reduce the infection rate. 
	The system phase on days 51-200 is a result of the different models such that:
	\begin{enumerate}
		\item Model 1: Days 51-200: no restrictions.
		\item Model 2: Days 51-200: cycles of one week with no restrictions and one week of full lockdown.
		\item Model 3: Days 51-200: cycles of one week with no restrictions and two weeks of full lockdown.
	\end{enumerate}

	Each model was tested with the following constraints:
	\begin{itemize}
		\item with and without moderate social distancing (SD) on days 51-200;
		\item with and without infection from unknown sources (infected people not traceable to any known infection chain);
		\item with and without strict SD for symptomatic patients after the 14th day (sick people with symptoms are forced to keep 8 meters from healthy people, equivalent to strict home isolation).
	\end{itemize}
	Hence, for each population density, $ N $, we have 24 different simulations that will use later to assess both SD and lockdown on the number of cases as a function of time. 
	\section{Results}
	
	\begin{figure}[h!]
		\includegraphics[width=1\linewidth]{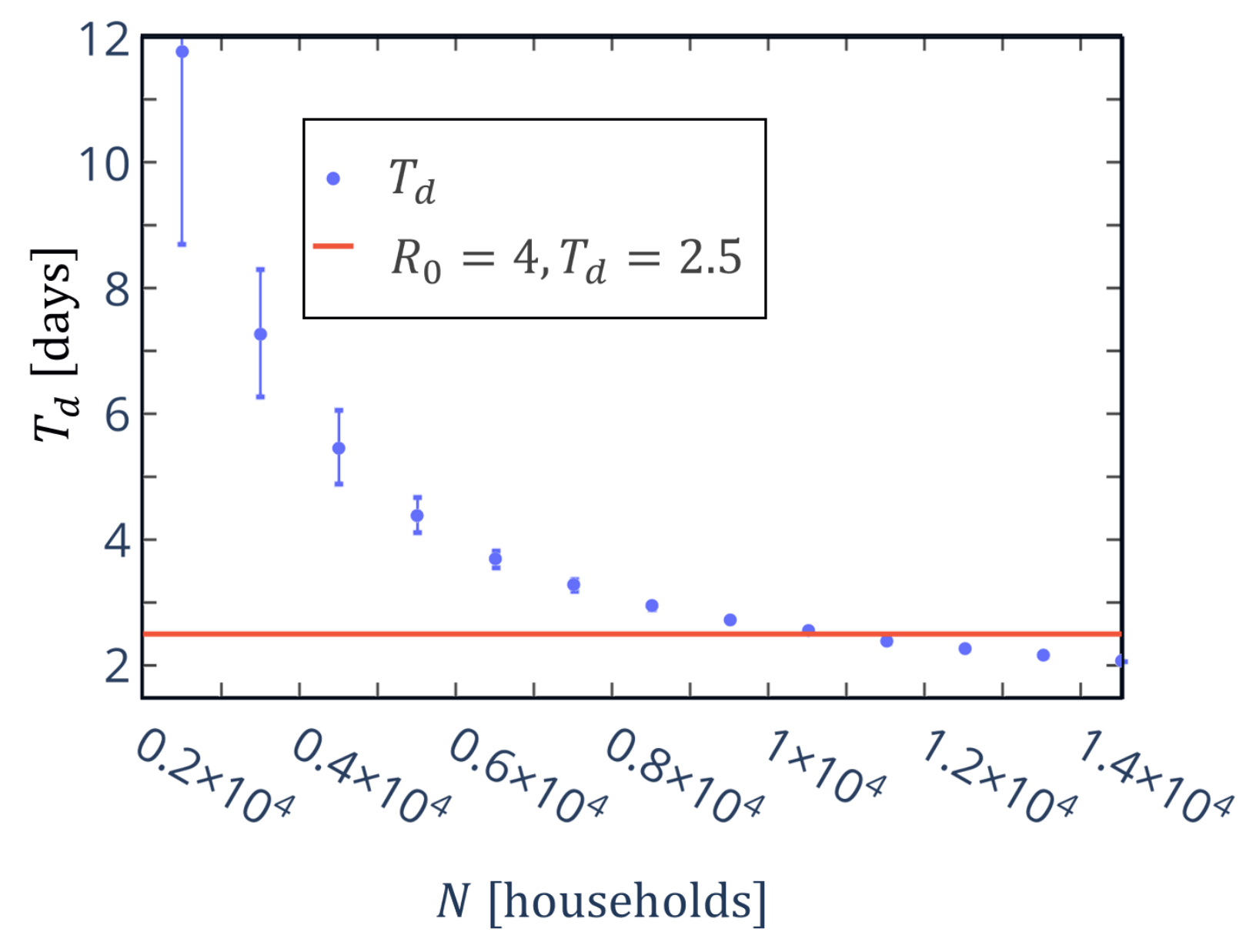}
		\caption{The doubling time, $T_d$ of the percentage of active cases from the total population in the first 14 days as the function of the population density, $N$ (dotes). The solid line marks the $T_d=2.5$ days value, corresponding to $R_0=4$. }
		\label{fig_Td}
	\end{figure} 
	
	For COVID-19 epidemics, the observed values of $R_0$, suggest that each infection directly generates $2-4$ more infections in the absence of countermeasures like social distancing \cite{karin2020adaptive,flaxman2020report}. The doubling time, $T_d$, is a function of $R_0$~\cite{visscher2020covid19}, 
	such that the higher $R_0$, the lower $T_d$. In particular, $T_d=2.5$ days corresponds to $R_0$=4.

	\cblack 
	In our model, both $R_0$ and $T_d$ are directly obtained from the simulation and not pre-assumed. Figure ~\ref{fig_Td} presents the doubling time, $T_d$ as a function of the effective density, $N$. We have calculated the percentage of active cases from the total population in the first 14 days for each $ N $. The calculation was performed in the heating phase only, since there are no restrictions. This allows for a sort of calibration of the model to encompass the intrinsic features of the disease.

	Figure~\ref{fig_Td} shows that $T_d$ is a decreasing function of the density. \cblack 	It was shown in Ref.~\cite{visscher2020covid19} that for low $R_0$ $(1<R_0<1.5)$, the doubling time spans between $12<T_d<20 $ days, in contrast to high $R_0 (3.5<R_0<4)$, where $T_d$ is$\approx 2.5$ days, where $\underset{N \to \infty}{\lim}T_d\approx1.5$ days. The significant difference between high and low $R_0$ is reflected in our calculations by a high error for high doubling times, which are compatible with low population densities. 
	
	%\cblack
	The relation between $T_d $ and $R_0$ as shown in Fig.~\ref{fig_Td} \cblack means that attempting to describe a wide area's current situation by means of some average value of $R_0$ might be highly unappropriated. On the other hand, the spread of the disease over smaller, more homogeneous areas in which the population shares a certain degree of mobility and social behavior would be quite well described by this parameter.

	To give an example, a value of $T_d\approx2.5$, corresponding to $R_0=4$, in the current model would correspond to $N=1.1\cdot10^4$ households per squared kilometer. While this might appear an unreasonably high density for an average urban context, it is still much lower than the average density in kindergartens, university classrooms, crowded social, religious or sports events, etc. Hence, for the next subsection, we will consider the value of $N=1.1\cdot10^4$ as representative of potentially dangerous situations present daily before the beginning of the pandemics. This will also show how different constraints affect the initial $R_0=4$, which is the highest estimation for $R_0$ without restrictions. 
	\begin{widetext}
		
		\begin{figure*}[h]
			\centering
			% \vspace{20pt}
			\includegraphics[width=\linewidth]{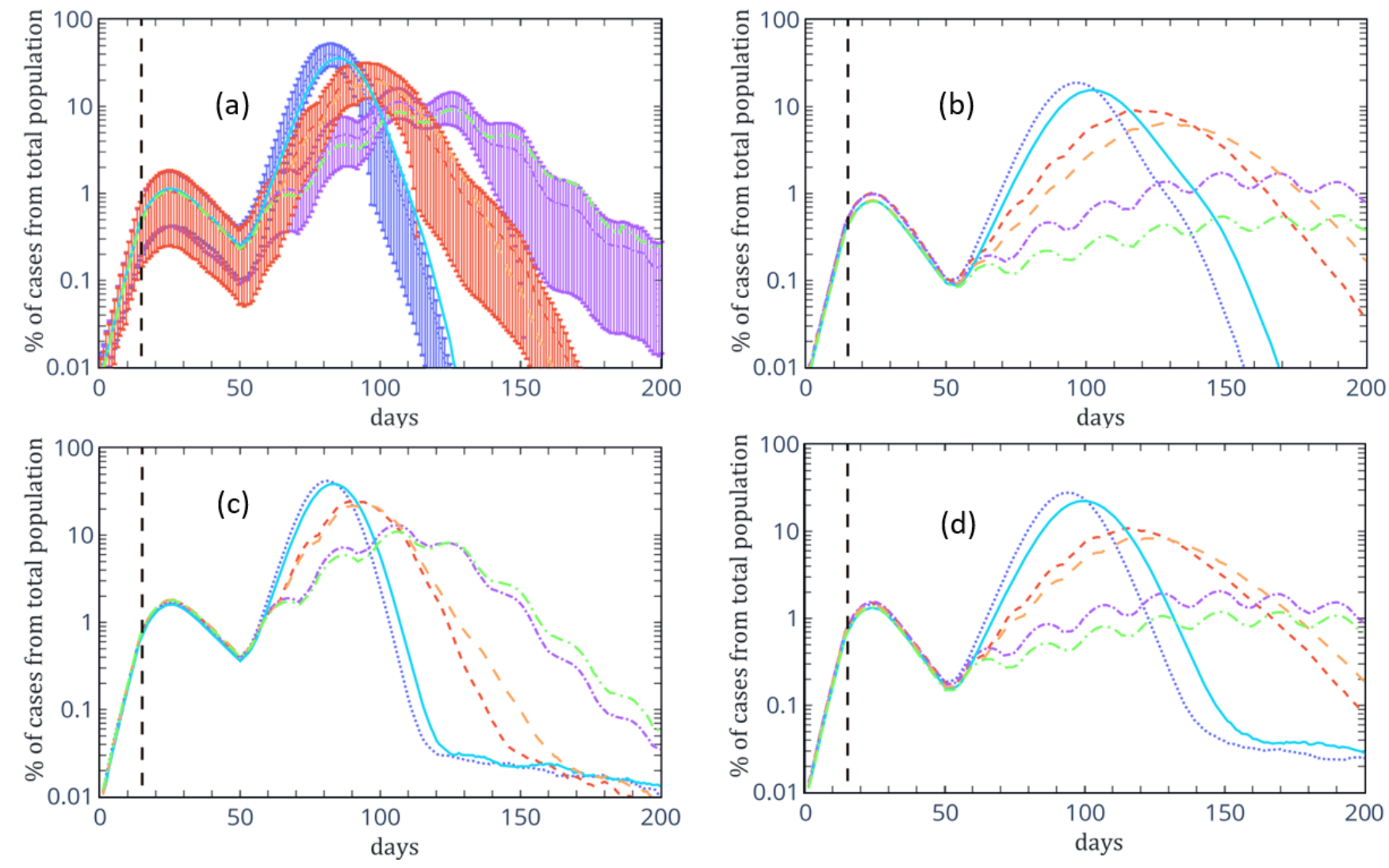}
			\vspace{8pt}
			\caption{Numerical results for the percentage of active cases from the total population for the case of infection.
				The upper panels (a) and (b) present the numerical result for infection without unknown sources, while the lower panels (c) and (d) show the numerical result for infection with unknown sources. For both cases, in the left panels (b and d), the numerical results are for the case that sick people with symptoms must keep 8 meters from healthy people (strict SD). For all panels, the solid (dotted) line is model 1 with (without) moderate social distancing (SD). The long (short) dashed line is model 2 with (without) moderate SD, while the long (short) dotted-dashed line is model 3 with (without) moderate SD. The dashed vertical line is located on day 15, the first day of the lockdown. Errorbars in panel (a) originate from the simulation's rmsd, computed from the 100 different samples generated to compute each curve. On the other curves, the uncertainty is similar and is omitted for improving the readability of the figures. }\label{fig_Y}
			\vspace{-20pt}
		\end{figure*}
	\end{widetext}
	
	The various probability densities are sampled by means of standard techniques, in the spirit of a kinetic Monte-Carlo simulation, for predicting the number of Coronavirus cases as a function of time for different lockdown models and external constraints. Each case has been run 100 times (Note that each run starts with the same initial condition, only one sick person. Since the Monte Carlo algorithm is based on random numbers, we expect that every run will yield slightly different results. Repeating the simulation for 100 separated runs evaluates the algorithm's robustness). Results have been averaged and analyzed to determine the statistical error. Figure~\ref{fig_Y} shows our numerical results for the different simulations for the case of $R_0=4$ in the first 14 days. The upper (lower) panels of Figure~\ref{fig_Y} are the numerical result without (with) unknown sources. In panels (b) and (d), the numerical results are for the case that sick people with symptoms must keep 8 meters from healthy people (strict SD).
	%\begin{widetext}
	
	%\end{widetext}
	
	As previously pointed out, the numerical results, presented in Figure~\ref{fig_Y}, do not make any assumption on the doubling time, $T_d$, but only on some observed features of the disease. Even though it is challenging to model each Coronavirus infected area's specific characters, some characteristics are common to all simulations.
	The first 50 days have the same constraints- no restriction from the first day up to day 14 and a full lockdown from day 15 until day 50. Our numerical results show that even for $R_0=4$, forcing a lockdown after 14 days from the first case, controls the spread of the Coronavirus in a way that the peak number of cases occurs on day 24, while the rate of increase in the number of cases starts decreasing as of the 15$^{\text{th}}$ day. Also, since the average time for recovery is 14 day (though up 40 days for very severe cases), and the typical infection period ranges from the 3$^{\text{rd}}$ day until the 7$^{\text{th}}$ day, the decreasing rate of the number of active cases during the lockdown is much slower than the increasing rate of the number of active cases without restrictions (as seen in various countries around the world\cite{endcoronavirus}). Hence, we find that for an initial heating period of 14 days, the necessary lockdown period (i.e., the cooling time) required to reduce the number of active cases is much longer than 14 days.

	As of today, many countries examine different exit strategies due to the decreased number of active cases. 
	In our simulations, we have tested several a few such exit strategies using different constraints. All of our numerical results indicate that the isolation of symptomatic patients (strict social distancing, panels (b) and (d) of Figure~\ref{fig_Y}) is effective and can reduce the peak number of active cases by about a factor two without further restrictions, and up to a factor of 10 for model 3. Also, from Figure~\ref{fig_Y}, we find that moderate social distancing can reduce the number of active cases, but never as effectively as home isolation of symptomatic patients.
	
	\begin{center}
		\begin{figure}[h]
			\includegraphics[width=\linewidth]{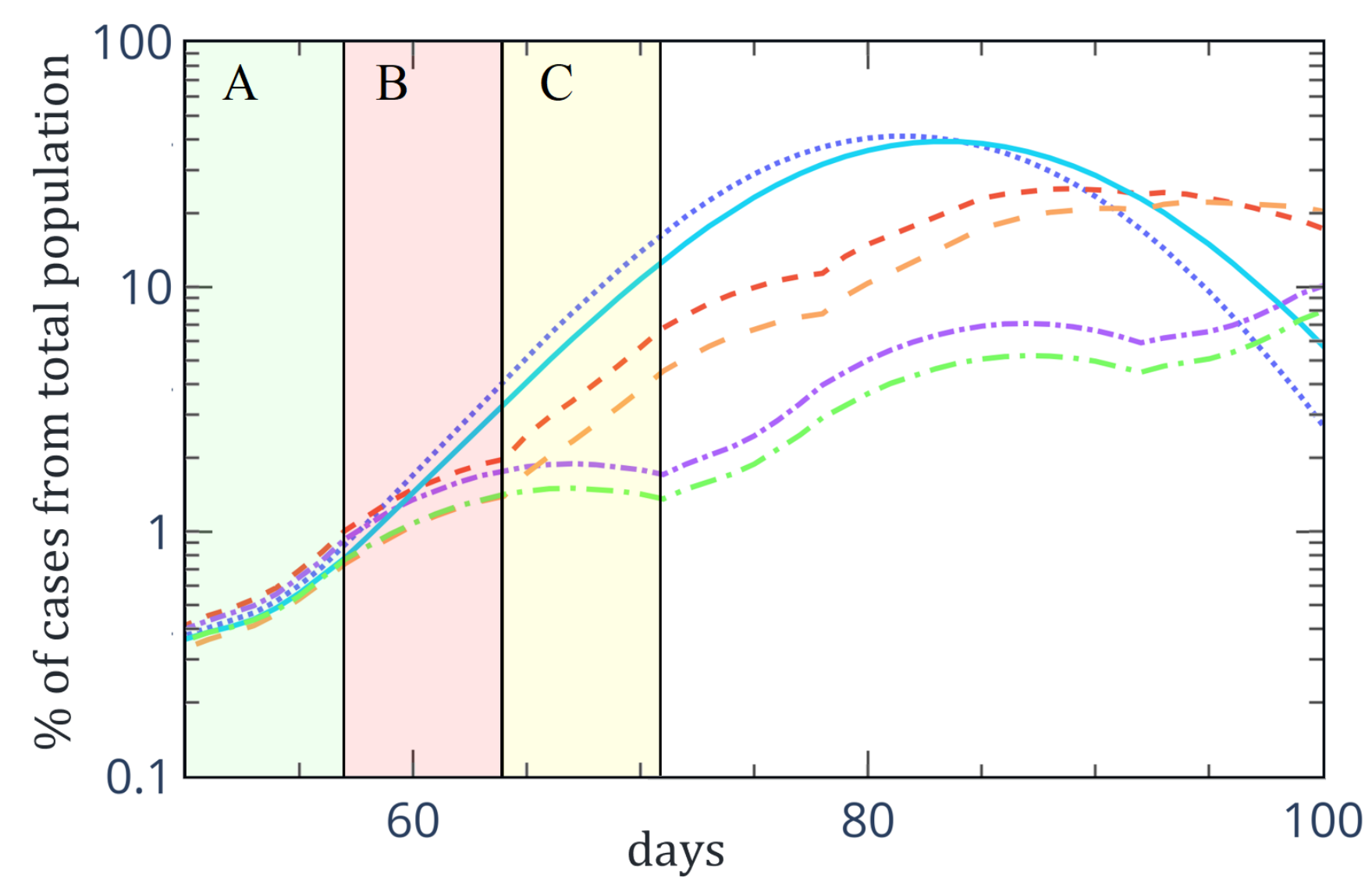}
			\caption{Inset of Figure~\ref{fig_Y} (c) for the case of infection with additional unknown sources. The solid (dotted) line is model 1 with (without) social distancing (SD). The long (short) dashed line is model 2 with (without) SD, while the long (short) dotted-dashed line is model 3 with (without) SD. }\label{fig_Y_zoom}
		\end{figure}
	\end{center}

	Figure \ref{fig_Y_zoom} is an inset of Figure~\ref{fig_Y} (c). From Figure~\ref{fig_Y_zoom} is it easy to see the effect of the cyclic no-restrictions/lockdown pattern. In principle, since the no-restriction/lockdown pattern is periodic in time, we would expect that the number of active cases as a function of time will have the same periodicity. For all the three models, there are no restrictions from day 51 until day 57 (green area, \cblack A\cblack ). From day 58 until day 64 (red area, \cblack B\cblack), we impose lockdown in models 2 and 3, which is reflected in a more moderate increase in the number of active cases than that of days 51 until 57. From day 65 until day 72 (yellow area, \cblack C\cblack), there is still full lockdown in model 3, while there are no restrictions in models 1 and 2. From Figure~\ref{fig_Y_zoom} it is easy to see that the increased effectiveness of model 3 in reducing the number of active cases originates from the fact that, since it takes more than a week to cool the system (which is a result of the infection period) , a week-week strategy cannot cause significant cooling of the system. An effective exit strategy might be based on time cycles, must include at least ten days of lockdown (see, for example, Ref~.\cite{karin2020adaptive}). Also, one can use different (no time based) no-restorations/lockdown patterns such that presented in Refs.~\cite{chandak2020epidemiologically,khadilkar2020optimising}. \cblack The lockdown strategies presented in Refs.~\cite{chandak2020epidemiologically,khadilkar2020optimising} are not time-dependent, in contract, they are based on optimization of a fixed number of infected/expired people over time.
	In general, our model can also be used for building these kinds of no-restrictions/lockdown pattern, for each population density. \cblack
	
	\begin{widetext}
		\begin{figure*}[]
			\vspace{-20pt}
			\includegraphics[width=1\linewidth]{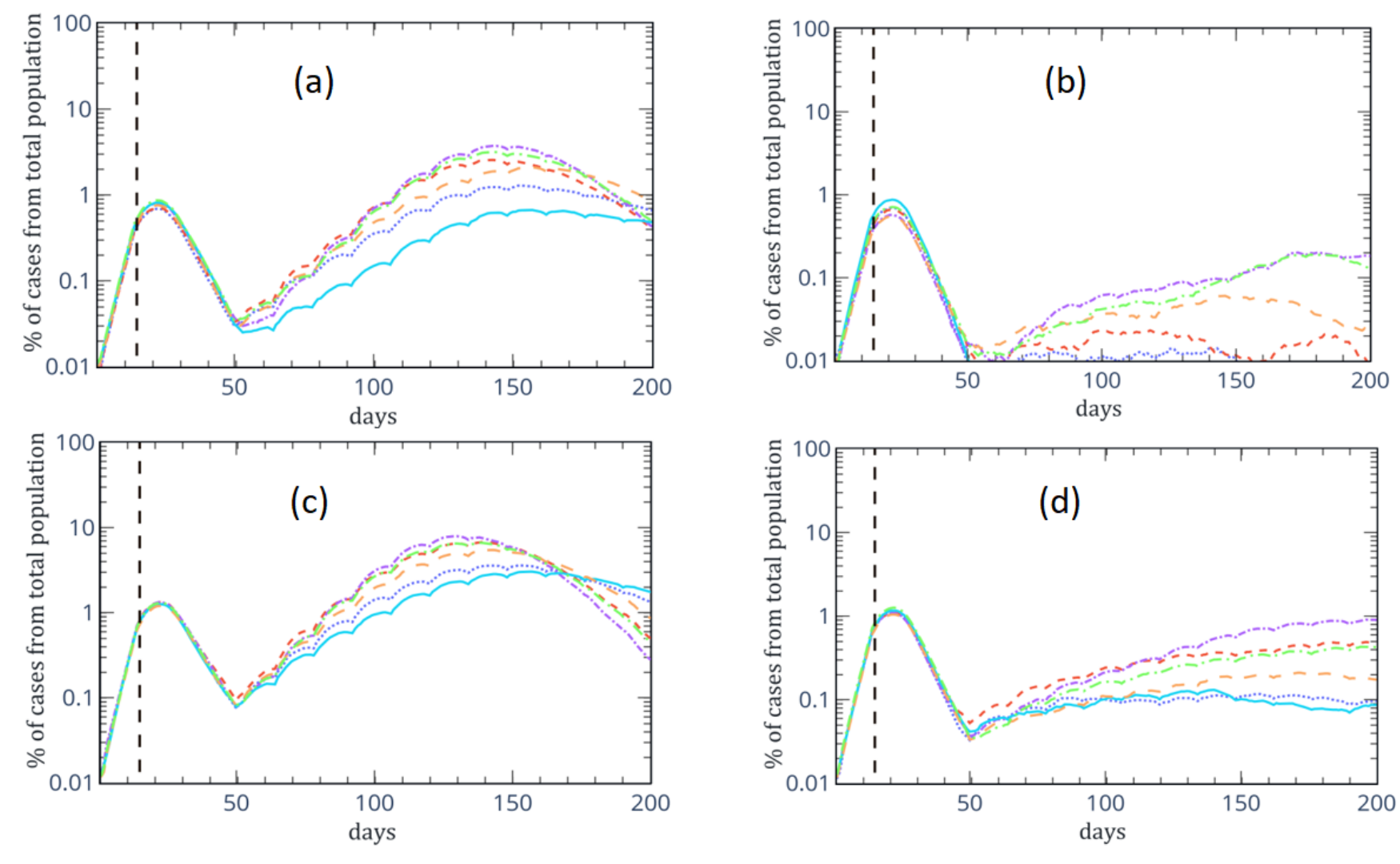}\\
			\caption{Numerical results for the percentage of active cases from the total population for the case of infection.
				The upper panels (a) and (b) present the numerical result for infection without unknown sources, while the lower panels (c) and (d) present the numerical result for infection with unknown sources. For both cases, in panels (b) and (d), the numerical results are for the case that sick people with symptoms must keep 8 meters from healthy people (strict SD).
				For all panels the solid (dotted) line is for $\sigma_{\text{low}}=0.5\times 10^{0.1}$ meters with (without) moderate social distancing (SD). The long (short) dashed line is for $\sigma_{\text{low}}=0.5\times 10^{0.3}$ meters with (without) moderate SD, while the long (short) dotted-dashed line is for $\sigma_{\text{low}}=0.5\times 10^{0.4}$ with (without) moderate SD. }\label{fig_Y2}
		\end{figure*}
	\end{widetext} 
	
	In Figure~\ref{fig_Y2} we present our numerical results for a 4/10 days cyclic exit strategy for different $\sigma_{\text{low}}$. Similar to Figure~\ref{fig_Y}, also here, the upper (lower) panels of Figure~\ref{fig_Y} are the numerical result without (with) unknown sources. For both cases, in the left panels (b and d), the numerical results are for the case that sick people with symptoms must keep 8 meters from healthy people (strict SD).
	%\begin{widetext}
	% \end{widetext}
	
	Figure \ref{fig_Y2} shows that, as predicted, for instance, in Ref.~\cite{karin2020adaptive}, a 4/10 days cyclic exit strategy is useful for controlling the infection rate accompanied by home isolation of symptomatic patients. The compression between Figure~\ref{fig_Y} and Figure~\ref{fig_Y2} indicates that although the maximal percentage of the active case is similar for both the week/two-weeks pattern and the 4/10 pattern, there are differences between the two patterns.
	For the 4/10 pattern, this cyclic pattern induces a moderate increase in the percentage of active cases but without a local decrease in the number of patients until the peak on the 130$^{th}$ day. In contrast, for the case of the week/two-weeks pattern, the two-weeks lockdown will cause a local decrease in the percentage of active cases and a more steep increase in the percentage of active cases during the no-restrictions week accomplished by a peak in the percentage of active cases on the 110$^{th}$ days. Hence, the compression between the two patterns implies that a cyclic no-restrictions/lockdown pattern can control the spread of the epidemic along with daily life, even for an initial doubling time of $2.5$ days.
	\cblack
	\section{Comparison with real data (Sweden)}
\begin{figure}[h!]
		\vspace{-20pt}
		\includegraphics[width=1\linewidth]{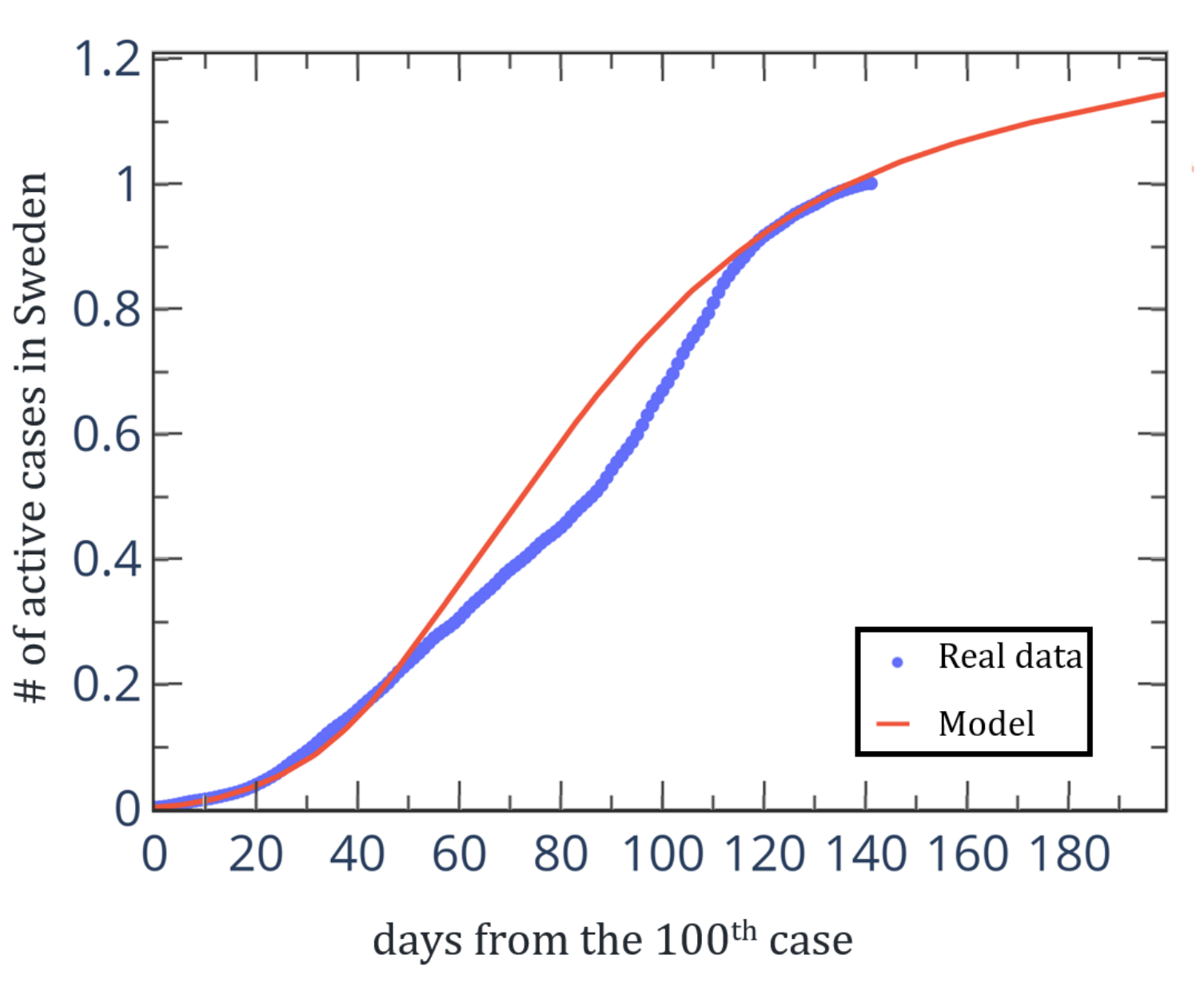}\\
		\caption{The total number of active cases in Sweden (normalized) from the 100$^{th}$ case until today (dotes). The solid line is our prediction for the spreading of the Coronavirus for a population density of $N=3500$ households.}\label{fig_Sweden}
	\end{figure}

	From the beginning of March 2020, Sweden takes a different approach from the rest of the world by not imposing a policy of lockdown on its citizens.
	Therefore, it is of interest to examine the \textbf{rate} of an increase in the total number of active cases in Sweden from the beginning of March till today\cite{endcoronavirus}, compared to our model under the assumption of no restrictions. In Figure \ref{fig_Sweden}, we show the total number of active cases in Sweden (normalized) from the 100$^{th}$ case until today, in comparison to our predictions for a population density of $N=3500$ households. Note that this population density is much more diluted than the population density that was used for the previous simulations, which may explain the slow rate of increase of virus spread, even when there are no limits. Figure \ref{fig_Sweden} shows that our model can predict the spread of the virus for different societies, reflected with varying densities of population. 
	
	\cblack
	\section{Conclusions}
	This paper presented a kinetic Monte-Carlo algorithm for modeling different scenarios of the infection rate of the novel Coronavirus disease. This model's main feature lies in its extreme flexibility and in the fact that the parameter $R_0$ is obtained from the simulation and not pre-assumed. It can rather be used, in principle, as a way to better tune up the other parameters based on the post-processing of clinical and epidemiological data.

	Although it is challenging to model the specific characters of each Coronavirus infected area, our results show that strict social distancing and a cyclic time pattern might help to keep the infection rate under control over a long period, even for an intrinsic doubling time of $2.5$ days and in the presence of infection from unknown sources. 
	Our ability to model and stand for differences between the different lockdown patterns sharpens the need for physical and mathematical models that allow examining different ways to reduce the spread of the epidemic. From the physical point of view, effective strategies for controlling the infection rate of a specific area should lower its effective temperature as much as possible by keeping social distancing and avoiding creating hot spots such as those related to high concentrations of people on a daily basis.

	\begin{acknowledgments}
		
		We thank R. Milo and his group at Weizmann Institute of Science for sharing their epidemiological data and for fruitful discussions. \end{acknowledgments}

	The data that support the findings of this study are available from the corresponding author upon reasonable request.
	
	\bibliography{references}
\end{document}